\documentstyle[12pt]{article}
\begin{document}
\title{REAL FORMS OF COMPLEX QUANTUM ANTI-DE-SITTER ALGEBRA
$U_{q}(Sp(4;C))$ AND THEIR CONTRACTION SCHEMES}
\author{{\em  Jerzy Lukierski}\thanks{Partially supported
by the Swiss National Science Foundation}\/ \thanks{On leave of
absence from the Institute for Theoretical Physics, University
of Wroc{\l}aw, ul. Cybulskiego 36, 50205 Wroc{\l}aw, Poland},
{\em Anatol Nowicki$^{*}$}\thanks{On leave of absence from
the Institute of Physics, Pedagogical University, Plac
S{\l}owia\'{n}ski 6, 65020 Zielona G\'{o}ra, Poland}
{\em and Henri Ruegg$^{*}$}
\\ D\'epartement de Physique Th\'eorique,\\
Universit\'e de Gen\`eve, 24 quai Ernest-Ansermet, \\
1211 Gen\`eve 4, Switzerland}
\date{}
\maketitle
\newcommand{\ostar}{O \kern -1.4ex {\raise -0.65ex \hbox{*}}}
\newcommand{\oostar}{O \kern -1.4ex {\raise -0.5ex \hbox{*}}}

\newcommand{\RRR}{R\hspace{-0.4cm}R}
\newcommand{\NNN}{N\hspace{-0.31cm}N}
\newcommand{\ZZZ}{Z\hspace{-0.4cm}Z}
\newcommand{\ds}{\displaystyle}
\newcommand{\un}{\underline}
\newcommand{\be}{\begin{equation}}
\newcommand{\ee}{\end{equation}}
\newcommand{\ba}{\begin{array}}
\newcommand{\ea}{\end{array}}
\newcommand{\bd}{\begin{description}}
\newcommand{\ed}{\end{description}}
\newcommand{\po}{Poincar\'{e}\/ \/}
\begin{abstract}
We describe four types of inner involutions of the Cartan-Weyl
basis providing  (for $ |q|=1$ and $q$ real) three types of
real quantum Lie algebras: $U_{q}(O(3,2))$
(quantum $D=4$ anti-de-Sitter), $U_{q}(O(4,1))$  (quantum
$D=4$ de-Sitter) and $U_{q}(O(5))$. We give also two types of
inner involutions of the Cartan-Chevalley basis of
$U_{q}(Sp(4;C))$ which can not be extended to inner
involutions of the Cartan-Weyl basis. We outline twelve contraction
schemes for quantum $D=4$ anti-de-Sitter algebra. All these
contractions provide four commuting translation generators,
but only two (one for $ |q|=1$, second for $q$ real) lead to
the quantum \po algebra with
an undeformed space rotations $O(3)$ subalgebra.
\end{abstract}
\newpage
\section{Introduction}
Recently the formalism of quantum groups and quantum Lie
algebras$^{1)}$ (see e.g. [1-6]) has been applied to physical
space-time symmetries. Several authors have looked for $q$-deformations
of $D=4$ Lorentz group and $D=4$ Lorentz algebra [7-11].
In other   papers the contraction schemes have been used
to obtain the quantum deformation of semisimple Lie
algebras describing Minkowski or Euclidean group
of motions. In particular there were obtained: \\
a) quantum deformations of $D=2$ and $D=3$ Euclidean
and Minkowski geometries, described as  quantum Lie
algebra or as quantum group [12,13] \\
b) quantum deformation of $D=4$ \po algebra [14] \\
Quite recently also several deformations of $D=2$ supersymmetry
algebra in its Euclidean [15-17] as well as Minkowski
[15,17] version were obtained.

In our recent paper [14] we obtained the quantum deformation of
$D=4$ \po algebra by contracting the Cartan-Weyl basis of
a particular real form $U_{q}(SO(3,2))$ ($ |q|=1$)
of the second rank quantum Lie algebra $U_{q}(Sp(4;C))$.
Performing these calculations we realized that the
chosen scheme is not unique. We found that there are at
least two questions which should be answered in detail
for the physical applications of any quantum Lie
algebra $U_{q}(\hat{g})$ of rank $\geq 2$: \\
a) The classification of involutions of the quantum
Lie algebra $(U_{q}(\hat{g}), \Delta, S, {\cal E})$, where
$\Delta$ denotes comultiplication
$(\Delta : U_{q}(\hat{g}) \to U_{q}(\hat{g})
\otimes U_{q}(\hat{g})), S$ - the antipode
("quantum inverse" ) ($S : U_{q}(\hat{g}) \to
U_{q}(\hat{g}))$ and $\epsilon$ is the counit
$(\epsilon : U_{q}(\hat{g}) \to C$)$^{2)}$. In principle one
can introduce four types of involutions (see Sect. 2).\\
b) The extension of inner involutions of Cartan-Chevalley
basis to the full Lie algebra basis, described by the
Cartan-Weyl generators. It appears that for physical
applications one should "solve"
the Serre relations, and study the reality condition
for the deformed Lie algebra generators.\\
In this paper we address these questions for the simple case
of rank $2$ quantum Lie algebra $U_{q}(Sp(4))$, and provide the
answers. It appears that different real structures imply
different signatures as well as different restrictions on
$q$ ($ |q|=1$ or $q$ real). In particular in the case of
$U_{q}(O(3,2))$ the Cartan subalgebra of $O(3,2)$ can be
chosen maximally compact $(O(2) \oplus O(2))$ or maximally
noncompact $(O(1,1) \oplus O(1,1))$, and two triples
$( e_{i}, e_{-i}, h_{i})$ ($i=1,2$) describing
Cartan-Chevalley basis of $Sp(4)$ for different real
structures describe different real forms of
$SL(2,C)$ ($O( 3)$ or $O(2,1)$). If we follow the method
ref. [14] and perform the contraction procedure for
different real structures we  obtain different deformations
of the four-dimensional inhomogeneous space-time algebras:
the quantum \po algebras with Minkowski signature (3,1) or the
ones
describing four-dimensional quantum geometry
with the signature (2,2). Further, the "physical"
quantum \po algebra (with signature (3,1)) can be
deformed in two different ways:

a) with undeformed $O(3)$ subalgebra of $D=3$ space rotations.

b) with undeformed $O(2,1)$ subalgebra.

The first case was obtained in our recent paper [14], with
$ |q|=1$. If we consider all possible real structures, both cases
a) and b) can be accompanied with the reality conditions
$ |q|=1$ as well as $q = {\bar{q}}$ ($q$ real). As a
consequence we conclude that in the deformation formulae
given in [14] for the quantum \po algebra also the purely
imaginary values of
deformation parameter
 $\kappa$ may appear. We obtain that modulo some
numerical coefficients, depending on the choice of the
involution, the $sin {P_{0} \over \kappa}$
and $cos \; {P_{0} \over \kappa}$ terms can be replaced
by $sinh \; {P_{0} \over \kappa}$ and
$cosh \; {P_{0} \over \kappa}$ terms, providing agreement
with earlier results for three - dimensional case
[12,13].

The plan of our paper is the following. In Sect. 2 we shall
rewrite the Hopf bialgebra $U_{q}(Sp(4;C))$ in Cartan-Weyl
basis, with comultiplication table, antipodes, counits
and Serre relations replaced by bilinear algebraic relations.
In Sect. 3 we shall consider two types of the Hopf algebra
involutions, satisfying respectively the conditions
$$
i)\qquad \qquad  S((S(a^{*}))^{*}) = a \Leftrightarrow
S \circ * = * \circ S^{-1}
\eqno(1a)
$$
$$
ii) \qquad \qquad S^{-1}((S(a^{\otimes}))^{\otimes}) = a
\Leftrightarrow S \circ \otimes = \otimes \circ S
\eqno(1b)
$$
We show that there are four different ways of introducing in total
16 real structures of Cartan-Weyl basis of $U_{q}(Sp(4))$.
(eight of type 1a) and eight of type 1b)). Expressing the
Cartan-Weyl generators of $U_{q}(Sp(4))$ as quantum deformation
of $O(3,2)$ Lie algebra basis, one can relate the choice of
real structures with the choice of signature of real
$D=5$ orthogonal algebras $O(5-k, k)$ ($k=0,1,2$). The results
are presented in Table 1. For completeness we shall present
also the involutions $*_{s} = S \circ *$
which due to the relation (1a) satisfy $*^{2}_{s}=1$.
They provide the examples of inner involutions of Cartan-Chevalley
basis which however cannot be extended to Cartan-Weyl basis.
In Sect. 4 we discuss the contraction schemes of different
real forms of $U_{q}(Sp(4))$, which are obtained for 16 choices
of real structures (12 for quantum anti-de-Sitter, 2 for
quantum de-Sitter, and 2 for $U_{q}(O(5))$, which can
be treated as quantum $D=4$ Euclidean de-Sitter algebra).
Finally in Sect. 5 we present general remarks and comments.
\section{$U_{q}(Sp(4))$ in Cartan-Weyl basis}
Let us recall that the Lie algebra $C_{2} \equiv Sp(4)$
can be described by the following standard choice of simple
roots (see e.g. [19])
$$
\alpha_{1} = \left( {1 \over \sqrt{2}}, - {1 \over \sqrt{2}} \right)
\qquad \alpha_{2} = (0, \sqrt{2} )
\eqno(2.1)
$$
which leads to the following symmetrized Cartan matrix
$\alpha_{ij} = <\alpha_{i}, \alpha_{j}>$
($i=1,2$):
$$
\alpha = \left(\matrix {1 & -1 \cr
-1 & 2}\right)
\eqno(2.2)
$$
The Drinfeld-Jimbo $q$-deformation of $U_{q}(Sp(4))$
is described by the following deformation of Cartan-Chevalley
basis
$(e_{i}, e_{-i}, h_{i})$, corresponding to simple roots
(2.1)
( $[x]\equiv (q - q^{-1} )^{-1} \cdot
(q^{x} - q^{-x} )$)
$$
\ba{c}
[e_{i}, e_{-j} ] = \delta_{ij} [h_{i}]_{q} \cr\cr
[h_{i}, e_{\pm j} ] =
\pm \alpha_{ij} e_{j} \qquad
[h_{i}, h_{j} ] = 0
\ea
$$
restricted also by the $q$-Serre relations
$$
\ba{c}
\left[ e_{\pm \alpha_{1}} \left[ e_{\pm \alpha_{1}}
\left[ e_{\pm \alpha_{1}}, e_{\pm \alpha_{2}} \right]
_{q^{\pm 1}} \right] _{q^{\pm 1}} \right]_{q^{\pm 1}} = 0\cr \cr
\left[e_{\pm \alpha_{2}} \left[ e_{\pm \alpha_{2}},
e_{\pm \alpha_{1}} \right]_{q^{\mp 1}} \right]_{q^{\mp 1}} = 0
\ea
\eqno (2.4)
$$
where $\left [e_{\alpha}, e_{\beta}\right]_{q} = e_{\alpha}e_{\beta} -
q^{-<\alpha,\beta>} e_{\beta}e_{\alpha}$
The coproduct and antipodes are given by the
formulae
$$
\ba{l}
\Delta (h_{i}) = h_{i} \otimes 1 + 1 \otimes h_{i} \cr \cr
\Delta  ( e_{\pm i} ) = e_{\pm i} \otimes
q^{ h_{i} \over 2} +
q^{- \;{h_{i} \over 2}} \otimes e_{\pm i}
\ea
\eqno(2.5)
$$
and
$$
S(h_{i}) = - h_{i} \qquad
S(e_{i\pm} ) = - q^{\pm {1 \over 2} d_{i}}
e_{i \pm}
\eqno(2.6)
$$
where $d_ {i} = <\alpha_{i}, \alpha_{i}>
= (1,2)$.  \\
In order to write the Cartan-Weyl basis one has
to introduce the generators corresponding to nonsimple
roots. $( \alpha_{1} + \alpha_{2},
2 \alpha_{1} + \alpha_{2} )$. Introducing the normal
order of roots $(\alpha_{1}, \alpha_{4}=
2 \alpha_{1} + \alpha_{2}$, $\alpha_{3} = \alpha_{1}
+ \alpha_{2}, \alpha_{2} )$ and using the following defining
relations for nonsimple  generators staying in
normal order $(\alpha, \alpha + \beta, \beta)$
[20,21]
$$
\ba{l}
e_{\alpha + \beta} = e_{\alpha} e_{\beta} -
q^{- <\alpha, \beta>} e_{\beta} e_{\alpha} \cr \cr
e_{-\alpha - \beta} = e_{-\beta} e_{-\alpha}
- q^{<\alpha, \beta>} e_ {-\alpha} e_{-\beta}
\ea
\eqno(2.7)
$$
one obtains ($[A,B]_{q} \equiv
AB - qBA$)
$$
\ba{ll}
e_{3} = [e_{1}, e_{2}]_{q}
\qquad
& e_{-3} = [e_{-2}, e_{-1} ]_{q^{-1}} \cr \cr
e_{4} = [e_{1}, e_{3} ]
\qquad
& e_{-4} = [e_{-3}, e_{-1} ]
\ea
\eqno(2.8)
$$
The complete set of commutation relations for
$U_{q}(C_{2})$ in $q$-deformed Cartan-Weyl basis
was given in [14] . We recall that
$$
\ba{ll}
[e_{i},e_{-j} ] = \delta_{ij} [h_{i}]_{q} & \cr \cr
[h_{i},h_{j} ]= 0 & \cr \cr
[h_{i},e_{\pm j} ] = \pm \alpha_{ij}e_{\pm j} &\cr \cr
[e_{3}, e_{-3}] = [h_{3}]_{q} \quad
&h_{3} = h_{1} + h_{2} \cr \cr
[e_{4}, e_{-4} ] = [h_{4}]_{q}
\quad
&h_{4} = h_{1} + h_{3}
\ea
\eqno(2.9a))
$$
The set of bilinear relations equivalent to q-Serre
relations takes the form:
$$
\ba{ll}
[e_{1}, e_{4}]_{q^{-1}} = 0,
&[e_{-2}, e_{-3} ]_{q} = 0 \cr \cr
[e_{4}, e_{3}]_{q^{-1}} = 0,
&[e_{-3}, e_{-4} ]_{q} = 0 \cr\cr
[e_{3}, e_{2} ]_{q^{-1}} = 0,
&[e_{-4}, e_{-1} ]_{q} = 0
\ea
\eqno(2.9b)
$$

The formulae for coproduct and antipode for the generators
(2.8) look as follows:
$$
\ba{l}
\Delta(e_{3})= e_{3} \otimes q^{{1 \over 2 }h_{3}}
+ q^{-{1\over 2}h_{3}} \otimes e_{3}
+ \left(q^{-1} - q\right)
q^{-{1 \over 2} h_{2}} e_{1} \otimes e_{2}
q^{{1 \over 2} h_{1}} \cr \cr
\Delta(e_{-3}) =
e_{-3} \otimes q^{{1 \over 2} h_{3}} +
q^{- {1 \over 2}h_{3}} \otimes
e_{-3} +
\left(q - q^{-1} \right)
q^{- {1 \over 2} h_{1}} e_{-2} \otimes
e_{-1} q^{{1 \over 2}h_{1}}
\ea
\eqno(2.10a)
$$\\
$$
\ba{l}
\Delta(e_{4})= e_{4} \otimes q^{{1 \over 2}h_{4}}
+ q^{- {1 \over 2}h_{4}} \otimes e_{4} +
\left(q - q^{-1} \right) \left\{ \left(
1 - q^{-1} \right)\right. \cr \cr
\left. \cdot q^{-{1 \over 2} h_{2}} e^{2}_{1} \otimes
e_{2} q^{h_{1}} - q^{-{1 \over 2} h_{3}}
e_{1} \otimes e_{3} q^{{1 \over 2} h_{1}} \right\}
\cr \cr
\Delta(e_{-4}) = e_{-4} \otimes
q^{{1 \over 2}h_{4}} + q^{- {1 \over 2} h_{4}}
\otimes e_{4} +
\left( q - q^{-1} \right) \left\{ \left(
q - 1\right) \right.
\cr \cr
\left. \cdot q^{-h_{1}} e_{-2} \otimes e^{2}_{-1} q^{{1 \over 2} h_{2}}
+ q^{- {1 \over 2} h_{1}}
e_{-3} \otimes
e_{-1} q^{{1 \over 2} h_{3}} \right\}
\ea
\eqno(2.10b)
$$\\
and
$$
\ba{l}
S(e_{3}) = - q^{1 \over 2} e_{3} + q^{1 \over 2}
(1 - q^{2})e_{1}e_{2} \cr \cr
S(e_{-3}) = - q^{- {1 \over 2}} e_{-3}
+
q^{-  {1 \over 2}} \left( 1 - q^{-2} \right)
e_{-2} e_{-1} \cr \cr
S(e_{4}) = - qe_{4}
+ \left(1 - q^{2}\right) \left\{ (q - 1)
e^{2}_{1} e_{2} + e_{1}e_{3} \right\} \cr \cr
S(e_{-4}) = - q^{-1} e_{-4}
+ \left( 1 - q^{-2} \right) \left\{ \left( q^{-1} - 1\right)
e_{-2} e^{2}_{-1} + e_{-3} e_{-1}
\right\}
\ea
\eqno(2.11)
$$
The description of quantum algebra in Cartan-Weyl basis is an
{\underline alternative} description to the standard one, given
by (2.3 - 2.6). There are
two advantages of this approach:\\
a) The nonlinear Serre constraints are replaces by bilinear relations
(see (2.9b)) \\
b) In the limit $q \to 1$ one obtains the ordinary Lie algebra
relations.

In comparison with [14] we shall propose here  more general
relations between the "root" generators
$e_{a}, e_{-a}, h_{i}$,
$(a = 1 \cdot ; i=1,2)$ and the "physical"
rotation generators $M_{AB} = - M_{BA}$
$(A,B=0,1,2,3,4,)$, allowing for the $q$-dependent
rescaling of the "root" generators
$$
e_{\pm a} \to E_{\pm a} = C_{\pm a}
(q) e_{\pm a}
\eqno(2.12)
$$
where $C_{\pm} (1) = 1$. In fact a nontrivial choice of the numerical
coefficients $C_{\pm a}$ is necessary if we wish to obtain
the reality conditions for $M_{AB}$ from the inner
involution formulae for the Cartan-Chevalley basis,
satisfying the condition (1.1a).
\section{Real forms of $U_{q}(Sp(4))$ }
Let us consider the Hopf algebra ${\cal A}$ over $C$
with comultiplication $\Delta$ and antipode $S$. We shall
distinguish the following four types of involutive
homomorphisms in ${\cal A}$:

i) The $+$ - involution, which is an anti-automorphism
in the algebra sector, and automorphism in the coalgebra
sector, i.e. $(a_{i} \in {\cal A})$:
$$
\left( a_{1} \cdot a_{2} \right)^{+}
= a^{+}_{2} a^{+}_{1}
\qquad
\left( \Delta (a)\right)^{+} = \Delta (a^{+} )
\eqno(3.1)
$$

ii) The $*$-involution, which is an automorphism in the algebra
sector, and an antiautomorphism in the coalgebra sector, i.e.
$$
\left( a_{1} \cdot a_{2} \right)^{*} =
a^{*}_{1} a^{*}_{2}
\qquad \left( \Delta (a)\right)^{*} =
\Delta'(a^{*})
\eqno(3.2)
$$
where
$\Delta' = \tau \Delta = R\Delta R^{-1}$ ($\tau$-flip
automorphism in ${\cal A} \otimes {\cal A})$.

We assume further that the bialgebra involutions i), ii) satisfy
the relation (1a).

iii) The $\oplus$- involution, which is an antiautomorphism
in both algebra and coalgebra sectors, i.e.
$$
\left( a_{1} \cdot a_{2} \right)^{\oplus} =
\left( a_{2}\right)^{\oplus} \left(a_{1}\right)^{\oplus}
\qquad
\left(\Delta(a)\right)^{\oplus} = \Delta'\left(
a^{\oplus} \right)
\eqno(3.3)
$$

iv) The $\ostar$ - involution, which is an automorphism
in both algebra and coalgebra sectors, i.e.
$$
\left( a_{1} \cdot a_{2} \right)^{\oostar}
= \left(a_{1}\right)^{\oostar}\cdot \left( a_{2} \right)^{\oostar}
\qquad \left( \Delta \left( a\right) \right)^{\oostar}
= \Delta \left( a^{\oostar} \right)
\eqno(3.4)
$$
We assume for the bialgebra involutions iii), iv) the
relation (1b). We would like to recall that
the standard real structure of the Hopf algebra
${\cal A}$ is given by the involution i)
(see e.g. [2]).

Now we shall present 16 involutions of $U_{q}(Sp(4))$ -four
in every category i) - iv), distinguished by two
parameters $\lambda$ and $\epsilon$ (we choose further
$\epsilon = \pm 1, \lambda = \pm 1)$ -
which are inner on Cartan - Weyl basis($k = q^{{1\over2}h}$):

i) $+$ - involutions $( |q|=1)$
$$
\ba{lcl}
k^{+}_{i} = k_{i}
&\Leftrightarrow
&h^{+}_{i} = - h_{i} \cr \cr
e^{+}_{\pm 1} = \lambda e_{\pm 1}
& &e^{+}_{\pm 2} = \epsilon e_{\pm 2} \cr \cr
e^{+}_{\pm 3} = - \lambda \epsilon
q^{\mp 1} e_{\pm 3}
& & e^{+}_{\pm 4} = \epsilon q^{\mp 1} e_{\pm 4}
\ea
\eqno(3.5)
$$
We define the rescaled generators
$E_{\pm 3}, E_{\pm 4}$ by introducing in the formula (2.12)
$$
C_{\pm 3} (q) = C_{\pm 4} (q) = q^{\mp {1 \over 2}}
\eqno(3.6)
$$
satisfying the relations
$$
E^{+}_{\pm 3} = - \lambda \epsilon E_{\pm 3}
\qquad
E^{+}_{\pm 4} = \epsilon E_{\pm 4}
$$
We define the rotation generators as follows:
$$
\ba{c}{
\ba{ll}
M_{12} = h_{1}
&M_{34} = {1 \over \sqrt{2}}
(E_{-3} - E_{3} ) \cr \cr
M_{23} = {1 \over \sqrt{2}} \left(
e_{1} + e_{-1} \right)
& M_{24} = - {1 \over 2} \left( e_{2} + e_{-2} +  E_{4} + E_{-4}
\right) \cr \cr
M_{31} = {1 \over \sqrt{2}} \left( e_{1} - e_{-1}
\right)
& M_{14} = - {1 \over 2} \left( e_{2} - e_{-2} - E_{4} + E_{-4}
\right)
\ea}\cr \cr
{\ba{c}
M_{04} = h_{3} \cr \cr
M_{03} = {1 \over \sqrt{2}}
\left(E_{3} + E_{-3} \right) \cr \cr
M_{02} = {1 \over 2}
\left( e_{2} - e_{-2} + E_{4} - E_{-4}
\right) \cr \cr
M_{01} = {1 \over 2}
\left( e_{2} + e_{-2} - E_{4} - E_{-4} \right)
\ea}
\ea
\eqno(3.8)
$$
giving for $\epsilon = \lambda = -1$ the condition
$$
M_{AB} = - M^{+}_{AB}
\eqno(3.9)
$$
with
$$
A,B = 0,1,2,3,4
$$
and for q = 1
$$
\left[ M_{AB}, M_{CD} \right] =
g_{BC} M_{AD} + g_{AD} M_{BC}
- g_{AC}M_{BD} - g_{BD}M_{AC}
\eqno(3.10)
$$
where
$$
\epsilon = \lambda = -1:
\qquad
g_{AB} = diag(- + - + +)
\eqno(3.11a)
$$

By proper choice of "i" factors in the formulae (3.8)
one can achieve the condition (3.9) also for
other values of $\epsilon$ and $\lambda$. One gets
then the relations (3.10) with the following set of
metrics:
$$
\lambda = 1, \;\; \epsilon = -1:
\qquad g_{AB} = diag(- + - - +)
\eqno(3.11b)
$$
$$
\lambda = -1, \;\; \epsilon = 1:
\qquad
g_{AB} = diag(+ +  - + - )
\eqno(3.11c)
$$
$$
\lambda = 1, \;\; \epsilon = 1:
\qquad
g_{AB} = diag
\left( + + - - - \right)
\eqno(3.11d)
$$

The involution (3.5) with
$\epsilon = \lambda = 1$
corresponds in dual picture to  the one proposed in [1] for the real form
$Sp_{q}(2n;R)$ of the complex quantum group
$Sp_{q} (2n;C)$.

ii) $*$ - involution ($q$ real)
$$
\ba{lcl}
k^{*}_{i} = k^{-1}_{i}
& \leftrightarrow
& h^{*}_{i} = - h_{i}
\cr \cr
e^{*}_{\pm 1} = \lambda e_{\mp 1}
\qquad
&& e^{*}_{\pm 2} = \epsilon e_{\mp 2} \cr \cr
e^{*}_{\pm 3} = - \lambda \epsilon
q^{\pm 1} e_{\mp 3}
\qquad
& & e^{*}_{\pm 4} =
\epsilon q^{\pm 1} e_{\mp 4}
\ea \eqno(3.12)
$$
If we rescale the generators $e_{\pm 3}, e_{\pm 4}$
using the scaling factor (3.6) and introduce in suitable way the
rotation generators (3.8) with proper "i-factors" , in order
 to satisfy the condition
$$
M_{AB} = M^{*}_{AB}
\eqno(3.13)
$$
one can show that the choices of $\lambda$
and $\epsilon$ in (3.12) implies the following
$D=5$ orthogonal metrics$^{3)}$
$$
\lambda = - 1,\; \epsilon = -1:
\qquad g_{AB} = diag( + + + + +)
\eqno(3.14a)
$$
$$
\lambda = 1,\; \epsilon = -1:
\qquad
g_{AB} = diag(+ + + - +)
\eqno(3.14b)
$$
$$
\lambda = - 1, \; \epsilon = 1:
\qquad
g_{AB} = diag(-  + + + -)
\eqno(3.14c)
$$
$$
\lambda = 1, \; \epsilon = 1:
\qquad
g_{AB} = diag(- + + - - )
\eqno(3.14d)
$$
We see therefore that we obtained for six involutions (3.11a-d)
and (3.14c-d) the $O(3,2)$
metric, for the choice (3.14b) - the $O(4,1)$ metric,
and for (3.14a) - the $O(5)$ metric. All eight involutions
(3.5) and (3.12) satisfy the condition (1.1a).

iii) $\oplus$ - involution $( |q| = 1)$
$$
\ba{ll}
k^{\oplus}_{i} = k^{-1}_{i}
\leftrightarrow
& h^{\oplus}_{i} = h_{i}
\cr \cr
e^{\oplus}_{\pm 1} = \lambda e_{\mp 1}
& e^{\oplus}_{\pm 2} = \epsilon e_{\mp 2}
\cr \cr
e^{\oplus}_{\pm 3} = \lambda \epsilon e_{\mp 3}
& e^{\oplus}_{\pm 4} = \epsilon e_{\mp 4}
\ea
\eqno(3.15)
$$

In the case of the involutions (3.15) the rescaling
of "root" generators is not needed. If one choose "i - factors"
in the formula (3.8) properly, the relations (3.15) will imply
the reality conditions
$$
M^{\oplus}_{AB} = - M_{AB}
\eqno(3.16)
$$
and the following choices of the metric in the formula
( 3.10):
$$
\lambda = 1, \epsilon = 1:
\qquad
g_{AB} = diag(+,+,+,+,+)
\eqno( 3.17a)
$$
$$
\lambda = 1, \epsilon = -1:
\qquad
 g_{AB} = diag(- , +, +, +, -)
\eqno(3.17b)
$$
$$
\lambda = - 1, \epsilon = 1:
\qquad
 g_{AB} = diag(- , -, -, +, -)
\eqno(3.17c)
$$
$$
\lambda = - 1, \epsilon = - 1:
\qquad
g_{AB} = diag(+, -, -, +, + )
\eqno(3.17d)
$$
We would like to mention here that in our previous
paper [14] we considered the contraction scheme
for the involution (3.17b); the involution
(3.17a) was also presented in [14].

iv) $\ostar$- involution $(q \mbox{real})$
$$
\ba{ll}
k^{\oostar}_{l} = k_{i}
\Leftrightarrow
& h^{\oostar}_{i} = h_{i} \cr \cr
e^{\ostar}_{\pm 1} = \lambda e_{\pm 1}
& e^{\oostar}_{\pm 2} = \epsilon e_{\pm 2} \cr \cr
e^{\oostar}_{\pm 3} = \lambda \epsilon e_{\pm 3}
& e^{\oostar}_{\pm 4} = \epsilon e_{\pm 4}
\ea
\eqno(3.18)
$$

The choice of the involutions (3.18) implies for real
generators.
$$
M^{\oostar}_{AB} = M_{AB}
\eqno(3.19)
$$
the following choices of metric:
$$
\ba{ll}
\lambda = 1, \; \epsilon = 1:
& g_ {AB} = diag(- + - + + )
\cr \cr
\lambda = 1, \; \epsilon = -1:
&g_{AB} = diag(+ + - + - )
\cr \cr
\lambda = - 1, \; \epsilon = 1:
& g_{AB} = diag(- + - - +)
\cr \cr
\lambda = - 1, \; \epsilon = - 1:
& g_{AB} = diag(+ + - - -)
\ea
\eqno(3.20)
$$
By considering the involutions (3.15) and (3.18)
we see again that we obtained six involutions providing
$O(3,2)$ metric, one - the $O(4,1)$ metric,
and one giving $O(5)$ metric. All eight involutions (3.15)
and (3.18) satisfy the condition (1.1b).

In order to complete the discussion of involutions for
$U_{q}(Sp(4))$ we should consider also two types
of involutions, obtained by the composition of antipode map
and the involutions i), ii). Using the formulae
for the antipode one gets the following two  new
involutions:

v) ${*}_{s} = S \cdot * $

Because the antipode $S$ is an antiautomorphism of
algebra as well as coalgebra, the involution
$*_{s}$ is a standard real structure of Hopf
bialgebra behaving as
the $+$ - involution (see i)), and satisfies the relations [1]:
$$
\ba{l}
\left( a_{1} a_{2} \right)^{*_{s}} =
\left( a_{2}\right)^{*_{s}} \cdot \left( a_{1}
\right)^{*_{s}} \cr \cr
\left( \Delta (a)\right)^{*_{s}} =
 \Delta \left( a^{*_{s}} \right)
\ea
\eqno(3.21)
$$
On $U_{q}(Sp(4))$ the relations (3.21) imply  that
$q$ should be real. Besides the involution
$*_{s}$ takes the following explicite form:
$$
\ba{lll}
k^{*_{s}}_{i} = k_{i} &\Leftrightarrow
 &\qquad h^{*_{s}}_{i} = h_{i} \cr \cr
e^{*_{s}}_{\pm 1} = - \lambda q^{\mp {1 \over 2}} e_{\mp 1}
& & \qquad e^{*_{s}}_{\pm 2} = - \epsilon
q^{\pm 1} e_{\mp 2} \cr \cr
e^{*_{s}}_{\pm 3} = - \lambda \epsilon
q^{\mp {1 \over 2}} \tilde{e}_{\mp 3}
& & \qquad e^{*_{s}}_{\pm 4} = \epsilon q^{\mp 1} \tilde{e}_{\mp 4}
\ea
\eqno(3.22)
$$
where $\tilde{e}_{\pm a} = \tau e_{\pm a}$
$(a=3,4; \tau  - \mbox{flip automorphism})$, i.e.
$$
\ba{ll}
\tilde{e}_{3} = \left[ e_{2}, e_{1} \right]_{q}
\qquad
&\tilde{e}_{4} = \left[\tilde{e}_{3}, e_{1}\right] \cr \cr
\tilde{e}_{-3} = \left[ e_{-1}, e_{-2}  \right]_{q^{-1}}
\qquad
& \tilde{e}_{-4} = \left[ e_{-1}, \tilde{e}_{-3} \right]
\ea
\eqno(3.23)
$$
We easily see from the formula (3.22) that the involution
$*_{s}$ {\un does not have an extension to the
Cartan-Chevalley basis}.

vi) $+_{s} = S \cdot +$

In similar way one can introduce for $ |q|=1$
behaving as the $*$-involution (see  ii)). It is given by the
following outer involution of the Cartan-Chevalley
basis:
$$
\ba{lll}
k^{+_{s}}_{i} = k^{-1}_{l}
&\Leftrightarrow
 & \qquad h_{i} = h^{+_{s}}_{i} \cr \cr
e^{+_{s}}_{\pm 1} =
- \lambda q^{\pm {1 \over 2}} e_{\pm 1}
& & \qquad e^{+_{s}}_{\pm 2} = - \epsilon
q^{\pm 1} e_{\pm 2} \cr \cr
e^{+_{s}}_{\pm 3} = - \lambda \epsilon
q^{\pm {1 \over 2}} \tilde{e}_{\pm 3}
& & \qquad e^{+_{s}} _{\pm 4} = \epsilon q^{\pm 1}
\tilde{e}_{\pm 4}
\ea
\eqno(3.24)
$$
\section{Different contraction schemes}
In our previous paper [14] we have chosen the
$\oplus$-involution (3.15) with the choice of
the parameters $\lambda = 1, \epsilon = - 1$
(see 3.17b), providing the real quantum Lie
algebra $U_{q}(O(3,2))$. In this paper we have
obtained twelve  real quantum Lie algebras
$U_{q} (O(3,2))$ (see table 1)
\begin{table}
\scriptsize
\begin{tabular}{|c|c|c|c|c|c|c|c|}
\hline\cr
1&2&3&4&5&6&7&8 \cr \hline
Root ma-
&Type
&algebra
&Co-
&Metrics
&Cartan
&Nondeformed
&$D=4$\cr
pings;
&of
&
&algebra
&($g_{00}, g_{11},g_{22},g_{33},g_{44}$)
&subalgebra
&subalgebra
&subalgebra
\cr
values of $q$
&involution
&
&
&
&
& $D=3$
&$(q=1)$
\cr
\hline
$\Delta_{\pm} \to \Delta_{\pm}$
& $+$
&anti-
& aut.
&$(+ + - - - )$
&$NC\oplus NC$
&$O(2,1)$
&$O(3,1)$
\cr
&
&aut.
&
&$(- + - - - )$
&$NC\oplus NC$
&$O(2,1)$
&$O(3,1)$
\cr
$ |q|=1$
&
&
&
&$(+ + - + - )$
&$NC\oplus NC$
&$O(2,1)$
&$O(2,2)$
\cr
&&&
&$(- + - + +  )$
&$NC\oplus NC$
&$O(2,1)$
&$O(2,2)$
\cr
\hline
$\Delta_{\pm} \to \Delta_{\mp} $
&
&
&anti-
&$(+ + + + +)$
&$C\oplus C$
&$O(3)$
&$O(4)$
\cr
$q$ real
&*
&aut.
&aut.
&$(+ + + - +)$
&$C \oplus C$
&$O(2,1)$
&$O(3,1)$
\cr
&&&
&$(- + + + -)$
&$C \oplus C$
&$O(3)$
&$O(3,1)$
\cr
&&&
&$(- + + - - )$
&$C\oplus C$
&$O(2,1)$
&$O(2,2)$
\cr
\hline
$\Delta_{\pm} \to \Delta_{\mp}$
&
&anti-
&anti-
&$(+ + + + + )$
&$C\oplus C$
&$O(3)$
&$O(4)$
\cr
$ |q|=1$
&$\oplus$
&aut.
&aut.
&$(+ + + - +)$
&$C\oplus C$
&$O(2,1)$
&$O(3,1)$
\cr
&&&
&$(+ - - - +)$
&$C \oplus C $
&$O(3)$
&$O(3,1)$
\cr
&&&
&$(+ - - + +)$
&$C \oplus C$
&$O(2,1)$
&$O(2,2)$
\cr
\hline
$\Delta_{\pm} \to \Delta_{\pm}$
&
&
&
&$(- + - + + )$
&$NC \oplus NC$
&$O(2,1)$
&$O(3,1)$
\cr
$ |q|$ real
& $\ostar$
&aut.
&aut.
&$(+ + - + - )$
&$NC \oplus NC$
&$O(2,1)$
&$O(2,2)$
\cr
&&&
&$(- + - - + )$
&$NC \oplus NC$
&$O(2,1)$
&$O(2,2)$
\cr
&&&
&$(+ + - - - )$
&$NC \oplus NC$
&$O(2,1)$
&$O(3,1)$
\cr
\hline
\end{tabular}
\caption{DESCRIPTION OF REAL STRUCTURES AND CONTRACTIONS
OF $U_{q}(Sp(4;C))$
(Aut. $\equiv$ automorphism, antiaut. $\equiv$  antiautomorphism,
$C$ - compact
abelian $(O(2))$, $NC$ - noncompact abelian
$(O(1,1)); \; \Delta_{+} (\Delta_{-})$ - set of
positive (negative) roots) }
\end{table}

In particular $U_{q}(Sp(4))$ as a second rank
quantum algebra contains two quantum
$U_{q}(Sl^{(i)}(2,C))$ subalgebras
$(i=1,2)$, with generators $(e_{i}, e_{-i}, h_{i})$.
Because we have chosen the assignment provided by the
formulae (3.8) we obtain that
$$
SL^{(1)} (2;C) = (M_{12}, M_{23}, M_{31})
\eqno(4.1a)
$$
$$
SL^{(2)}( 2;C) =
\left(M_{04}, {1 \over \sqrt{2}}
\left(M_{02} + M_{14} \right),
{1 \over \sqrt{2}} \left(M_{01} - M_{24} \right) \right)
\eqno(4.1b)
$$
Our contraction is defined by rescaling the generators
$(\mu,\nu =1,2,3,4)$ [14]
$$
{\cal P}_{\mu} = {1 \over R}
M_{\mu 0}
\qquad M_{\mu \nu}\;\; \mbox{unchanged}
\eqno(4.2)
$$
the rescaling of the deformation parameter [12-14]
$$
ln q=
\cases{ {1 \over \kappa R}
&for $q$ real \cr \cr
{i \over \kappa R} &for $ |q|=1$ }
\eqno(4.3)
$$
and performing the limit $R \to \infty$. Following [14]
we obtain in this limit that

i) The quantum subalgebra $U_{q}(SL^{(1)}(2,C))$ becomes
a classical one, i.e.
$$
U_{q}(SL^{(1)}(2;C))\;\;{\lower 0.5em
\hbox{$ {\stackrel{\longrightarrow}
{\scriptstyle {R \to \infty}}}$}}\;\; U(SL^{1}(2;C)
\eqno(4.4)
$$

ii) The generators $P_{\mu} = lim_{R \to \infty}
{\cal P}_{\mu}$ commute
$$
\left[ P_{\mu}, P_{\nu} \right] = 0
\eqno(4.5)
$$
however their primitive coproduct is modified

iii) If we put all $P_{\mu}=0$, we obtain the quantum
deformation of six generators, describing the quantum
counterpart of Lorentz algebra. In table 1 we list these
algebras in column 8 in the free limit
$\kappa \to \infty$.

The choice of a real form for $Sp(4)$   leads to the choice
of a real form of $SL^{(1)}(2,C)$. These real three - dimensional
algebras are listed in table 1 in column 7, and due to the relation (4.4)
they are not deformed.

We see from table 1 that out of 12 real quantum algebras
$U_{q}(O(3,2))$ only two seem to be physically interesting
- one considered in [14] and second one given by $*$ - involution
(3.12) with $\lambda = - 1$ and $\epsilon = 1$. Because the
rescaling (3.6) for $R \to \infty$ disappears,
in the contraction limit we obtain

a) The algebra considered in [14], with the generators
satisfying the reality condition corresponding to the
notion of Hermitean operators in quantum theory

b) The algebra from [14] with imaginary parameter $\kappa$
 and the reality
condition for generators corresponding to the notion of real
number (or real function) in classical theory.
In such a case we obtain the cos and sin deformation terms
replaced by cosh and sinh terms.

We would like to recall that the conventional \po algebra is
obtained in the limit $\kappa \to \infty$.
\section{Final Remarks}
This paper has been written in a search for other contractions
describing quantum deformation of \po algebra. We keep
the contraction scheme the same as proposed
in [14] but we
apply it to
all possible real forms of
quantum de-Sitter algebra
$U_{q}(Sp(4))$.
 As a result we obtain two cases which are
 interesting as candidates for  quantum
 deformation of the \po algebra:
 one corresponding to quantum, and second
to classical realization of quantum algebra.

In our paper we restricted ourselves to the involutions satisfying
the conditions (1a-b) - describing two types of real structure
of Hopf bialgebra over $C$. If these conditions are not satisfied,
we shall still get real bialgebras. It is possible that other
physically interesting contractions are of this type.
Also it should be added that we did not realize
the involutions as adjoint operators on the representation
spaces of quantum algebra. In general one can say only
that the involutions which are algebra automorphisms are
related with classical  realizations, (involution = complex
conjugation), and the involutive antiautomorphisms
are good candidates for the adjoint operation defined for
quantum realizations (involution = Hermitean conjugation).

Finally, we would like to stress that the contraction produces
the deformation described by a dimensionfull parameter, in
our case mass-like parameter $\kappa$. It would be extremely
interesting to find the role in real world for such
a mass-like parameter. One of possible applications is
the regularization of local fields by
the introduction of
 large but finite value of $\kappa$. Unfortunately, in
order to make such an
idea more concrete  firstly the $\kappa$ - deformation  of Minkowski
geometry  should be formulated.

Acknowlegments

One of us (J.L.) would like to thank Dr. V. Dobrev for discussions. Two
of the authors (J.L.) and (A.N.) would like to thank the University of
Gen\`eve for its warm hospitality.

\section*{FOOTNOTES:}
1. Following [1] we distinguish here the quantum
group $G_{q}$ as the $q$-deformation of the (algebra of
functions on the) Lie group, and the quantum Lie
algebra $U_{q}(\hat{g})$, described by the
$q$-deformation of the universal enveloping algebra
$U(\hat{g})$ of the Lie algebra.\\
2. For the discussion of real forms of the complex quantum
Lie algebra $U_{q}(SL(2;c))$ see e.g. [18];
for discussion of real forms of quantum groups see [1]. In these
papers only the antiautomorphisms of algebra and automorphisms
of coalgebra are listed.\\
3. We recall that $*$-involution is a conjugation of
algebra, not changing the order in a product of
operators, and the condition (3.13) implies the reality
of the metric in (3.10).

\end{document}